\documentstyle[jltp,11pt,twoside,epsf,rotate,bookmath]{article}

\newcommand{\onlinecite}{\citen}
\newcommand{\AB}{{\small\it AB}}
\newcommand{\SBF}{{\small\it SBF}}
\newcommand{\ESG}{{\small\it ESG}}
\renewcommand{\AE}{{\small\it AE}}
\newcommand{\AFM}{{\small AFM}}
\title{
\vspace*{-40mm}{\small
\mbox{}\centerline{
Los Alamos Report: LA-UR-98-3139;\
submitted to Journal of Low Temperature Physics}}\\
\vspace*{35mm}
Thermal Conductivity of the Accidental 
Degeneracy and Enlarged Symmetry Group
Models for Superconducting UPt$_3$
}
\author{M.\ J.\ Graf$^{1,2}$, S.-K.\ Yip$^{1}$ and J.\ A.\ Sauls$^{1}$}
\address{
  $^{1}$Department of Physics \& Astronomy,
  Northwestern University,\\
  Evanston, Illinois 60208-3112\\[1ex]
  $^{2}$Center for Materials Science,
  Los Alamos National Laboratory,\\
  Los Alamos, New Mexico 87545 \\[2ex]
  {\rm (submitted in July 1998)}
}

\runninghead{M. J. Graf, S.-K. Yip and J. A. Sauls}{Thermal Conductivity 
of the Accidental Degeneracy Models \dots}
\begin{document}
\begin{abstract}
We present theoretical calculations of the thermal conductivity for
the ``accidental degeneracy'' and ``enlarged symmetry group'' models
that have been proposed to explain the phase diagram of UPt$_3$.
The order parameters for these models
possess point nodes or cross nodes, reflecting the broken symmetries of
the ground state. These broken symmetries lead to robust predictions
for the ratio of the low-temperature thermal conductivity for heat flow
along the $\hat{c}$ axis and in the basal plane. The anisotropy of
the heat current response at low temperatures is determined by the
phase space for scattering by impurities.
The measured anisotropy ratio, $\kappa_c/\kappa_b$, provides a strong
constraint of theoretical models for the ground state order parameter.
The accidental degeneracy
and enlarged symmetry group models based on no spin-orbit coupling
do not account for the thermal conductivity of UPt$_3$.
The models for the order parameter that fit the
experimental data for the $\hat{c}$ and $\hat{b}$ directions 
of the heat current are the {\it 2D}
E$_{1g}$ and E$_{2u}$ models, for which the order parameters possess 
line nodes in the $ab$-plane and point nodes along the $\hat{c}$ axis,
and the A$_{1g}$$\oplus$E$_{1g}$ model of Zhitomirsky and Ueda. This model
spontaneously breaks rotational symmetry in the $ab$-plane below $T_{c2}$
and predicts a large anisotropy for the $ab$-plane heat current.
\end{abstract}
\maketitle
\newpage

The observations of multiple superconducting
phases\cite{mul87,qia87,bru90,ade90} of UPt$_3$ have led to several
Ginzburg-Landau models based on different symmetry groups or symmetry
breaking scenarios in order to account for the phase diagram. In this
paper we show that these models for the phase diagram, which are based
on different pairing symmetries, lead to qualitatively different
predictions for the thermal conductivity at very low temperatures.

The pairing models of UPt$_3$ which account for the main features of
the phase diagram may be grouped into three classes. One class of
models is based on a two-dimensional representation of $D_{6h}$ with
the multicomponent superconducting order parameter coupled to a {\it
S}ymmetry-{\it B}reaking-{\it F}ield ({\it SBF}). The leading
candidates for the pairing state are the even-parity E$_{1g}$ model and
the odd-parity E$_{2u}$ model. Both models require a symmetry breaking
field in order to split the transition in zero field and produce
multiple superconducting phases in a field. The \SBF~ is generally
assumed to be the in-plane \AFM~ order parameter that onsets at
$T_N\simeq 5\,K$;\cite{hay92} however, other explanations 
of the \SBF~ have been suggested.\cite{mid93,min93,ell95}

Several authors have argued that the
apparent tetracritical point in the H-T phase diagram for field
orientations out of the basal plane is incompatible with the {\it 2D} E-rep
models.\cite{luk91,che93,mac91,mac93} This led to explanations of the phase
diagram in terms of multiple order parameters that are unrelated by
symmetry (\eg `\AB-models'),\cite{luk91,che93} as well as further
examination of the E-rep models.\cite{sau94,par96} The four possible E-rep
models are not equivalent; weak hexagonal anisotropy, as is reflected
by the weak in-plane anisotropy of $H_{c2}$,\cite{shi86b,kel94} allows
for an apparent tetracritical point for all field orientations provided
the order parameter belongs to an E$_{2}$ orbital
representation.\cite{sau94} The odd-parity, 
E$_{2u}$ representation with strong spin-orbit locking
of the order parameter with $\vd ||\hat{\vc}$ ($\vd$ is the
quantization axis along which the pairs have zero spin projection,
i.e. $\vd\cdot\vS=0$) also accounts for the anisotropic paramagnetic
limiting of $H_{c2}$ observed at low temperatures.\cite{shi86b,cho91}
The spin-singlet E$_{1g}$ model appears to be incompatible with
both the tetracritical point for $\vH \not\perp \hat{\vc}$ and the
anisotropic Pauli limit for $H_{c2}$. However, Park and Joynt\cite{par96}
argue that there is enough freedom in the E$_{1g}$
model to account for all existing experimental data. Analysis of 
current heat transport data at low temperatures is accounted for equally
well by either E-rep. We suggest below that ultra-low temperature
measurements of the thermal conductivity for varying impurity concentrations
can also differentiate between the two E-rep models.
Both E-rep models have recently been challenged by
the observation of a nearly temperature independent Knight shift for
$\vH || \hat{\vc}$.\cite{tou98} Tou and co-workers\cite{tou98}
favor the \ESG-model of Machida \et\cite{mac93} based on
spin-triplet pairing with weak spin-orbit coupling.
Thus, our current understanding of the spin structure and parity of the 
order parameter for UPt$_3$ is unclear because there is conflicting
information from the NMR measurements of the spin susceptibility and
measurements of the anisotropic paramagnetic limit of the
upper critical field. However, as we show below, the low temperature
thermal conductivity of UPt$_3$ is in strong conflict with the non-unitary
spin-triplet model of Machida \et\cite{mac93}

The \AB-models are based on an assumed near accidental degeneracy of two
different pairing channels which are unrelated by
symmetry.\cite{luk91,che93} In these models the phase diagram is
accounted for by {\it two} primary order parameters belonging to
different irreducible representations of $D_{6h}$, which are selected
in order to enforce a tetracritical point in the GL theory for the H-T
phase diagram. The accidental degeneracy models which have been
investigated in most detail assume that the two order parameters belong
to different one-dimensional representations of the same parity, \eg
A$_{2u}$ and B$_{1u}$.\cite{che93} Another group of models for the
phase diagram may be described as {\it E}nlarged-{\it S}ymmetry-{\it
G}roup models. These are hybrids of the accidental degeneracy
and \SBF~ models. The \ESG-models assume an accidental degeneracy in the
form of a ``hidden symmetry'', which is lifted by a weak \SBF~ coupling,
similar to that of the E-rep models. Two different \ESG-models have 
been proposed.\cite{mac93,zhi93,zhi96}
The model of Zhitomirsky and Ueda\cite{zhi96}
is based on an enlarged {\it orbital} symmetry group
for UPt$_3$. In this model the primary pairing channel is assumed to be
the $d$-wave ($\ell=2$) channel of the full rotation group, and the
degeneracy of the $d$-wave manifold is lifted by weak
hexagonal anisotropy:  $\Gamma_2[SO(3)]\rightarrow A_{1g}\oplus
E_{1g}\oplus E_{2g}$. The A$_{1g}$ and E$_{1g}$ channels are assumed to
be weakly split, in order to account for the double transition.
Another formulation of an \ESG-model assumes spin-triplet pairing with
independent spin and orbital rotation symmetry, \ie {\it no} spin-orbit
coupling.\cite{mac93} In the model by Machida \et\cite{mac93} a
one-dimensional, odd-parity orbital representation is chosen in order
to enforce a tetracritical point for all field orientations. The
multicomponent order parameter corresponds to the three spin-triplet
amplitudes described by the $\vd$ vector, and the \SBF~ is assumed to
be the \AFM~ order parameter. The main features of each of these models
of the phase diagram and thermodynamic quantities of UPt$_3$ are
discussed in several articles.\cite{mac93,sau94,hef96,par96}

In this paper we investigate the low-temperature transport properties
for the ground states of the \AB~ and \ESG~ models. The
heat transport coefficients are sensitive to the nodal structure
and symmetry of the order parameter; at low temperatures the
lower-dimensional regions of the Fermi surface associated with the
nodal regions dominate the electronic heat
transport. The nodes of $\Delta(\vp_f)$ reflect the broken symmetries
of the superconducting phase. The most specific information
regarding the symmetry of the order parameter is reflected in the
ultra-low temperature limit of the components of the thermal
conductivity tensor, \ie $\lim_{T\rightarrow 0}\vkappa/T$.
The connection between the
symmetry of the pairing state and the low temperature thermal
conductivity is via impurity-induced Andreev scattering. Impurities in
unconventional superconductors lead to a band of low-energy Fermion
states which are formed by impurity-induced Andreev scattering and
the momentum-space structure of the order parameter.\cite{sau96}
In a dilute alloy a distribution of impurities leads to
a finite density of states, which we call an {\it Andreev band}.
The bandwidth and
density of states can be calculated from the self-consistent
$\hat{t}$-matrix and the leading order impurity 
self-energy terms.\cite{hir86,cho87,gra96a}
Impurity scattering within the Andreev band is then the dominant
heat transport mechanism at very low temperatures, \ie
$k_BT<\gamma$, where $\gamma$ is the width of the Andreev band. The
limiting values for $\vkappa/T$ are sensitive to the symmetry of
the order parameter, the geometry and dimensionality of the
nodes.\cite{gra96a,gra96b}

In an earlier paper \cite{gra96b} we compared the electronic thermal
conductivity measurements on UPt$_3$ by Lussier \et \cite{lus96b}
with our calculations based on the E-rep models. The experimental
fact that $\kappa_c(T)/\kappa_c(T_c) < \kappa_b(T)/\kappa_b(T_c)$ for
temperatures down to $T\simeq 50\,\mbox{mK}$ provides a strong constraint 
on models for the ground state order parameter and low-energy
excitation spectrum. We found
excellent agreement between theory and experiment for both the E$_{1g}$
and E$_{2u}$ representation [see also
Norman and Hirschfeld\cite{nor96} and
Lussier \et\cite{lus96b}]. The quality of our fits for both E-rep
models is evident in the temperature dependence of the normalized
anisotropy ratio
$[\kappa_c(T)/\kappa_b(T)]/(\kappa_c(T_c)/\kappa_b(T_c)]$ (Figs. 2 and
3 of Ref. \onlinecite{gra96b}).
From the published measurements, which extend down to
$T_c/10$, it is not possible to distinguish between the low temperature
phase of an even parity, spin singlet state (E$_{1g}$) and an odd
parity, spin triplet pairing state (E$_{2u}$). Further experimental
studies of the dependence of the anisotropy ratio on the impurity
concentration are needed to distinguish between these two E-rep models.
The differences in
symmetry between the E$_{1g}$ and E$_{2u}$ models may be tested at
ultra-low temperatures on nearly clean materials.
One possibility for distinguishing between the E$_{1g}$ and E$_{2u}$
ground states using transport measurements is based on the impurity scattering
dependence of the asymptotic values of $\vkappa/T$ at temperatures below
the impurity bandwidth, $k_BT \ll\gamma$. The predictions for the
anisotropy ratio $\cR=\lim_{T\rightarrow 0}
[\kappa_c(T)/\kappa_b(T)]/[\kappa_c(T_c)/\kappa_b(T_c)]$
as a function of the normal-state 
impurity scattering rate are shown in Fig. 1
for the two E-rep models.
The key feature is the universal (non-universal)
limit for the E$_{2u}$ (E$_{1g}$) model. 

\vspace{0.1in}
\noindent
\begin{minipage}{\textwidth}
\centerline{ \epsfysize=0.8\hsize\rotate[r]{\epsfbox{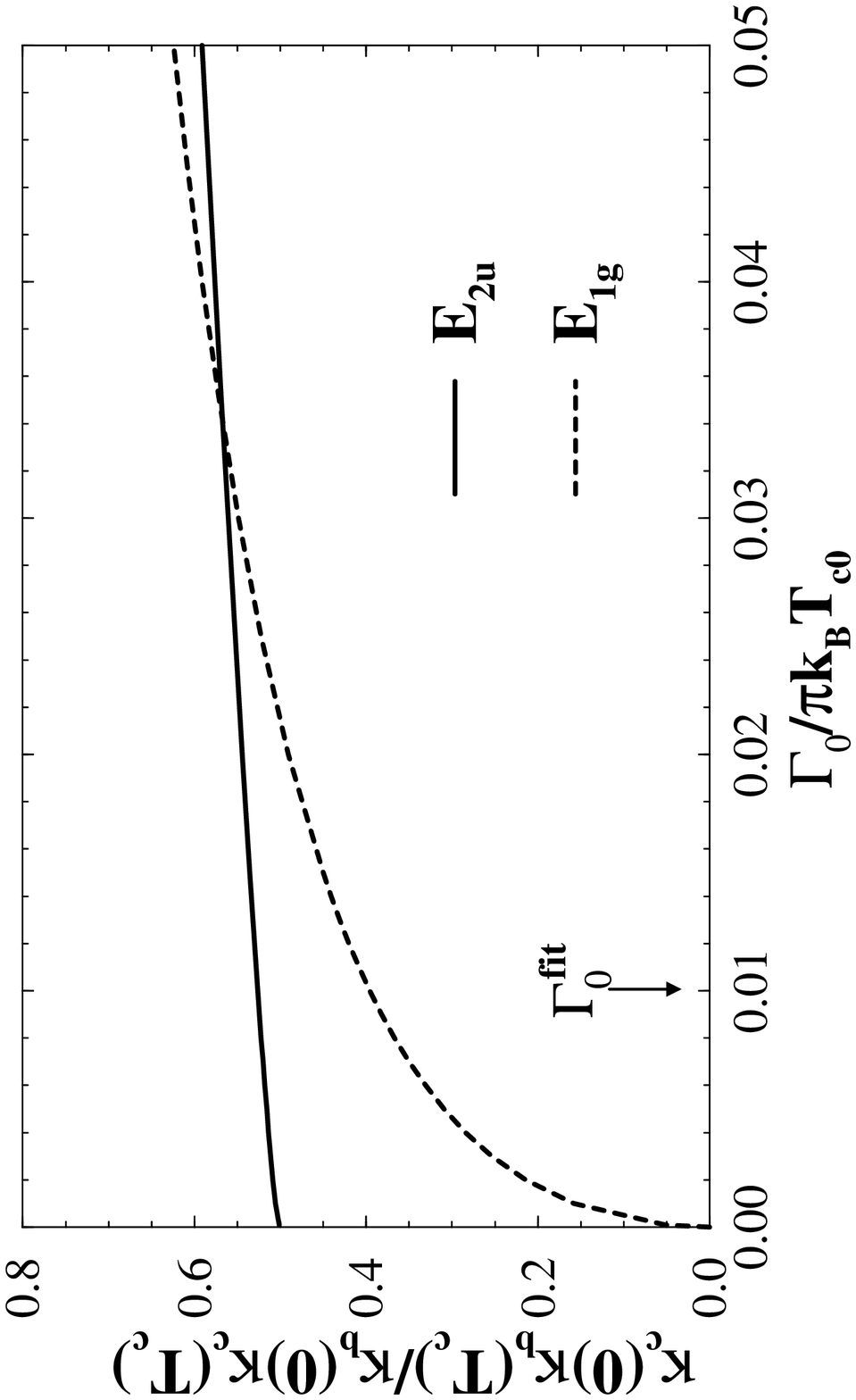} } }
\end{minipage}

\begin{center}
\begin{minipage}{1.0\textwidth}
{\small Fig.\ 1. The asymptotic low temperature anisotropy ratio
of the thermal conductivities for 
an E$_{1g}$ and E$_{2u}$ order parameter with the `best fit' parameters 
specified in Ref.~(\onlinecite{gra96b}).  The arrow indicates 
the estimated scattering rate obtained from the data of 
Ref.~(\onlinecite{lus96b}) for the fit to the E$_{2u}$ model.}
\end{minipage}
\end{center}
\vspace{0.1in}

Below we present our analysis of the electronic
thermal conductivity for the representative ground states of the \AB~ and
\ESG-models of UPt$_3$. We discuss the variational
forms for the order parameter
of these models, and derive asymptotic limits for the low
temperature thermal conductivity.
We compare our analytical results for the
asymptotic limits of $\vkappa/T$ with numerical calculations of
the thermal conductivity over the full temperature range below $T_c$,
and we compare the thermal conductivity
calculated within the \AB~ and \ESG~models with the experimental data on
UPt$_3$.\cite{lus96b,sud97} None of the \AB~ or \ESG-models based on
spin-triplet pairing without spin-orbit coupling account for
the temperature dependence and anisotropy of the thermal conductivity 
at low temperatures.
The only models which can account for the thermal conductivity data are
the {\it 2D} E$_{1g}$ and E$_{2u}$ models, and the orbital \ESG-model of 
Zhitomirsky and Ueda.\cite{zhi96} The \AE-model predicts a large $ab$-plane
anisotropy, which differentiates it from the 
E-rep models. We discuss the conditions under which the \AE-model can
account for the heat transport data in UPt$_3$.

\section*{Thermal Conductivity at Low Temperature}

Our analysis focuses on the low temperature region, $T\alt 0.5\,T_c$,
where heat transport is limited by scattering of electronic
quasiparticles off impurities.  We consider $s$-wave scattering from a
random distribution of defects in the strong scattering limit
($\delta_0\rightarrow \pi/2$). The assumption of nearly resonant scattering is
in agreement with transport measurements in the normal Fermi-liquid and
superconducting phases.\cite{lus96b} Inelastic ``electron-electron''
scattering is included by extrapolating the experimentally determined
normal-state inelastic scattering rate, $1/\tau_{ee}=A\,T^2$, to temperatures
below $T_c$; inelastic scattering is negligible below $T\simeq 0.3T_c$.

The transport properties of unconventional superconductors are more
strongly influenced by scattering from non-magnetic impurities than 
in conventional superconductors. One of the more striking
effects is the appearance of a band of low energy excitations deep in
the superconducting state.\cite{buc81,hir86,cho87,pre94a,bal95} This
occurs for non-magnetic impurities in unconventional superconductors
when the order parameter
changes sign around the Fermi surface. Impurity states develop
from the combined effects of impurity scattering and Andreev
scattering. In an unconventional superconductor with 
$\langle\Delta(\vp_{\!f})\rangle = 0$ and line or cross nodes,
a finite concentration of impurities leads to Andreev
states with a bandwidth, $\gamma \ll \Delta_0$, below which the density
of states is non-zero and almost constant for 
$\epsilon < \gamma$.\cite{gor85,cho88} This novel metallic band, deep in the
superconducting phase, can exhibit universal values for the
transport coefficients, \ie independent of the scattering rate at very
low temperatures, $k_B T\ll\gamma$.\cite{lee93,gra95,sun95,gra96a}

In Ref.~[\onlinecite{gra96a}] it was shown that the principal
components of the thermal conductivity tensor can be expressed in terms
of an {\it effective} transport scattering time that incorporates particle-hole
coherence of the superconducting state. For $T\rightarrow 0$,
\begin{eqnarray}
\lim_{T\rightarrow 0}\kappa_{i}(T) &=&
     \frac{v_{\!f,i}^2}{3}\,\gamma_S\,T \; \tau^{\tiny eff}_{i}\, \quad (i=a,b,c)\,,
\end{eqnarray}
where $v_{\!f,i}^2$ is the Fermi-surface average of
$[v_{\!f,i}(\vp_{\!f})]^2$, $\gamma_S=\frac{2}{3}\pi^2 k^2_B N_{\!f}$
is the normal-state Sommerfeld coefficient, and 
\begin{eqnarray}\label{lifetime}
\tau^{\tiny eff}_{i} = \frac{3\hbar}{2 v_{\!f,i}^2} \int d\vp_{\!f}\,
{ [v_{\!f, i}(\vp_{\!f})]^2\,
  \gamma^2
\over
  \left[ \Delta(\vp_{\!f})^2 + \gamma^2 \right]^{3\over 2}
} \,
\end{eqnarray}
is the {\it effective} transport scattering time.
The bandwidth of the Andreev states is given by
\begin{equation}
\gamma = \Gamma_u
   { \left< { \gamma\, (\Delta(\vp_{\!f})^2 + \gamma^2 )^{-1/2}  } \right>
    \over \cot^2 \delta_0 +  \left<
     { \gamma\, (\Delta(\vp_{\!f})^2 + \gamma^2 )^{-1/2}  }
   \right>^2 } \,,
\label{gamma}
\end{equation}
with the Fermi surface average $\langle \dots \rangle$, and
$\Gamma_u=n_i/(\pi N_{\!f})$. For a given impurity concentration,
$n_i$, this bandwidth is largest in the limit of unitarity scattering,
and depends on the order parameter symmetry through the structure and
geometry of the nodes of $\Delta(\vp_f)$ on the Fermi surface.

\section*{Order Parameter Models and Low Temperature Asymptotics}

Earlier experimental\cite{ shi86a,bro90a} and
theoretical\cite{miy86,sch86,hir86,mon86,pet86} transport studies
indicated that the excitation gap in UPt$_3$ has line nodes and probably
polar point nodes. This is reinforced by the thermal conductivity data
of Lussier \et\cite{lus96b} which is fit almost perfectly down to
$T\simeq 50\,\mbox{mK}$ by either the
E$_{1g}$ or the E$_{2u}$ models, both of which have a line node in the
$ab$-plane and polar point nodes. The differences in symmetry between the
E$_{1g}$ and E$_{2u}$ states are predicted to show up in the impurity
dependences of the anisotropy ratio $\lim_{T\rightarrow 0}\kappa_c/\kappa_b$ 
(see Fig. 1).\cite{gra96b}
However, as we show below the ground-state of the \AE-model has cross nodes
out of the basal plane located at polar angles, $\theta=\cos^{-1}(\pm 1/\sqrt{3})$.
The \AE-model can also account for the thermal conductivity data along
the $c$ and $b$ axes. The model also predicts a large anisotropy in the
$ab$-plane.

\vspace{0.3in}
{\small \noindent Table~1.\
Low temperature phases of several order parameter models.  The first two
entries are based on strong spin-orbit coupling and the symmetry group
$[D_{6h}]_{\tiny spin-orbit} \times {\cal T} \times U(1)$.  The third entry 
is based on no spin-orbit coupling and the \ESG-subclass 
$SU(2)_{\tiny spin} \times
[D_{6h}]_{\tiny orbit} \times {\cal T} \times U(1)$, and the last entry
belongs to the mixed symmetry orbital subclass of the \ESG,
$SO(3)_{\tiny spin-orbit} \times {\cal T} \times U(1)$.}
\begin{flushleft}
\renewcommand{\arraystretch}{1.2}
\begin{tabular}{ccccc}
\noalign{\smallskip}\hline\hline
    {$\Gamma$}
  & {${\cal Y}_\Gamma$}
  & {point}
  & {line}
  & {cross}
\\[1.0ex]
\noalign{\smallskip}\hline\noalign{\smallskip}
    $\rm E_{1g}$ & $z(x + i y)$
  & $\vartheta=0,\pi$	& $\vartheta=\frac{\pi}{2}$		& --
\\[1.0ex]
    $\rm E_{2u}$ & ${\bf \hat c} z (x + i y)^2$
  & $\vartheta=0,\pi$	& $\vartheta=\frac{\pi}{2}$		& --
\\[1.0ex]
    $\rm B_{1u}$ & $\vd\,\Im (x + i y)^3$
  & $\vartheta_m=m \pi$	 & $\varphi_n=n\,\frac{\pi}{3},$ & $\vartheta_m
    \wedge \varphi_n$
\\
  & &  $m=0, 1$ &  $n=0,..,5$	&
\\[1.0ex]
    $\rm A_{2u} \oplus B_{1u}$
  & $\quad {\bf \hat c}[{\rm A}\,z\Im (x + i y)^6$	& 
    $\vartheta_m=m \pi$ & $\varphi_n=n\,\frac{\pi}{3}$, & 
    $\vartheta_m \wedge \varphi_n$
\\
	& $\quad + {\it i}\,{\rm B}\,\Im (x + i y)^3 ]$ 
    &  $m=0, 1$ &  $n=0,..,5$	& 
\\[1.0ex]
    $\rm A_{1g} \oplus E_{1g}$
  & ${\rm A}\, (2 z^2 - x^2 - y^2)$ & -- & --
  & $\vartheta=\cos^{-1} \frac{\pm 1}{\sqrt{3}}$
\\
  & $+\quad {\it i}\,{\rm E} \, y z \quad$ & & & 
    $\wedge\, \varphi=0, {\pi}$
\\
\noalign{\smallskip}\hline\hline
\end{tabular}
\renewcommand{\arraystretch}{1.0}
\end{flushleft}
\vspace{0.3in}

Generic basis functions\cite{yip93c} for E-rep models, \AB~ and
\ESG-models for the ground state of UPt$_3$ are listed in
Table~1. All of these models have both line nodes
and point nodes - or ``cross nodes'' which are point nodes formed at
the intersection of line nodes in the two different basis functions of
the accidental degeneracy models. Order parameter models like the polar state
(A$_{1u}$), or combinations of the polar state with a nodeless A$_{1g}$
state are excluded by existing transport data.

Crystal symmetry alone does not determine the detailed structure of the
order parameter or excitation gap as a function of position on the
Fermi surface. In particular, the behavior of $\Delta(\vp_f)$ in the
vicinity of the nodes depends on material parameters. These `nodal'
parameters determine the rate at which the excitation gap opens near a
linear point or line node (\eg E$_{1g}$ ground state),
the curvature of a quadratic
point node (\eg E$_{2u}$ ground state) or the third derivative of a cubic point node
(\eg B$_{1u}$ state). In principle these nodal parameters can be calculated
by solving for the eigenfunctions of the linearized gap equation, but
this requires precise knowledge of the pairing interaction for momenta
on the Fermi surface. Our approach is to enforce the symmetry of $\Delta(\vp_f)$
through the positions of the nodes and to model the nodal structure
with phenomenological parameters which we fix by comparison with experiment.
Variational basis functions are constructed from the generic basis functions by
multiplying by a function, ${\cal F}_\Gamma$, which is invariant under all
crystal group operations.
This procedure allows us to introduce the Fermi surface 
anisotropy into the basis functions and independently model
the nodal regions of the excitation spectrum for line and point nodes.
Although the specific choice of functions ${\cal F}_\Gamma$ 
for a given representation is not unique, once we fix
the nodal parameters the low temperature thermal conductvity
is determined. As we have shown previously,\cite{gra96a,gra96b}
the low-temperature response is dominated by the
lower dimensional regions of the Fermi surface near 
the nodes of the order parameter; in these regions impurity induced
Andreev states with energies $\epsilon < \gamma \ll \Delta_0$ form 
a `metallic' band that determines the transport properties
at $k_B T < \gamma$.

A key feature required to account for heat transport
in UPt$_3$ is the existence of gapless excitations with both
$\vv_f\cdot\hat{\vc}\ne 0$ and $\vv_f\cdot\hat{\vb}\ne 0$.
Machida \et\cite{mac93} proposed an \ESG-model 
based on no spin-orbit coupling
and a 3D order parameter in spin space,
corresponding to the three possible
spin-triplet pair states, all of which
have the same orbital pair wave function.
They assume a 1D orbital function with A$_{2u}$ or
B$_{1u}$ symmetry.\cite{mac93} The corresponding
orbital basis functions are
\be
\Delta^{\tiny A_{2u}}(\vp_{\!f}) = 
\left[
  \Delta_0^A(T)\,\frac{343\sqrt{7}}{216}
  \hat p_z {\rm Im\,} (\hat p_{\!x} \!+\! i \hat p_{\!y})^6
\right]
  \; {\cal F}_{\tiny A_{2u}}(\vp_f; \mu_,\mu_{6})
\,,
\ee
\be
\Delta^{\tiny B_{1u}}(\vp_{\!f}) = 
\left[
\Delta_0^B(T)\,
  {\rm Im\,} (\hat p_{\!x} \!+\! i\hat  p_{\!y})^3 
\right] 
  \; {\cal F}_{\tiny B_{1u}}(\vp_f; \mu_,\mu_{3})
\,,
\ee
with $\Delta(\theta) \approx \Delta_0 \mu_n \theta^n$
near the point node, and
$\Delta(\theta) \approx \Delta_0 \mu \phi$
near the line node at $\phi=0$.
The B$_{1u}$ function has a cubic point and three {\it axial} line nodes.
The distinctive feature of this model 
is the large density of low-energy excitations near the poles.
The A$_{2u}$ function has a sixth-order point node and six
axial line nodes. Among the \AB-models, the A$_{2u}\oplus$B$_{1u}$ 
order parameter, $\Delta^{\tiny AB}(\vp_{\!f}) =
\Delta^{\tiny A_{2u}}(\vp_{\!f}) 
+ i \Delta^{\tiny B_{1u}}(\vp_{\!f})$,
has the same nodal features 
as the B$_{1u}$ model, \ie a cubic
point node and three axial line nodes.

The density of impurity-induced Andreev states
near the point and line nodes, 
and the heat current carried by these states,
depends sensitively on 
the gap function near the point or line nodes. For a cubic
point node at the poles,
$\Delta(\phi\!=\!\pi/6,\theta) \approx \mu_3 \Delta_0 \theta^3$,
which leads to $\kappa_c/T \sim \mu_3^{-2/3}$ as $T\to 0$.\cite{gra96a}
Thus, increasing $\mu_3$
tends to reduce the phase space for scattering among states 
near the cubic point node (see Fig.~2).
This is the case for both B$_{1u}$
and A$_{2u}\oplus$B$_{1u}$ models provided $\mu_3 < 1$.

\vspace{0.1in}
\noindent
\begin{minipage}{\textwidth}
\centerline{\epsfysize=0.5\hsize\rotate[r]{ \epsfbox{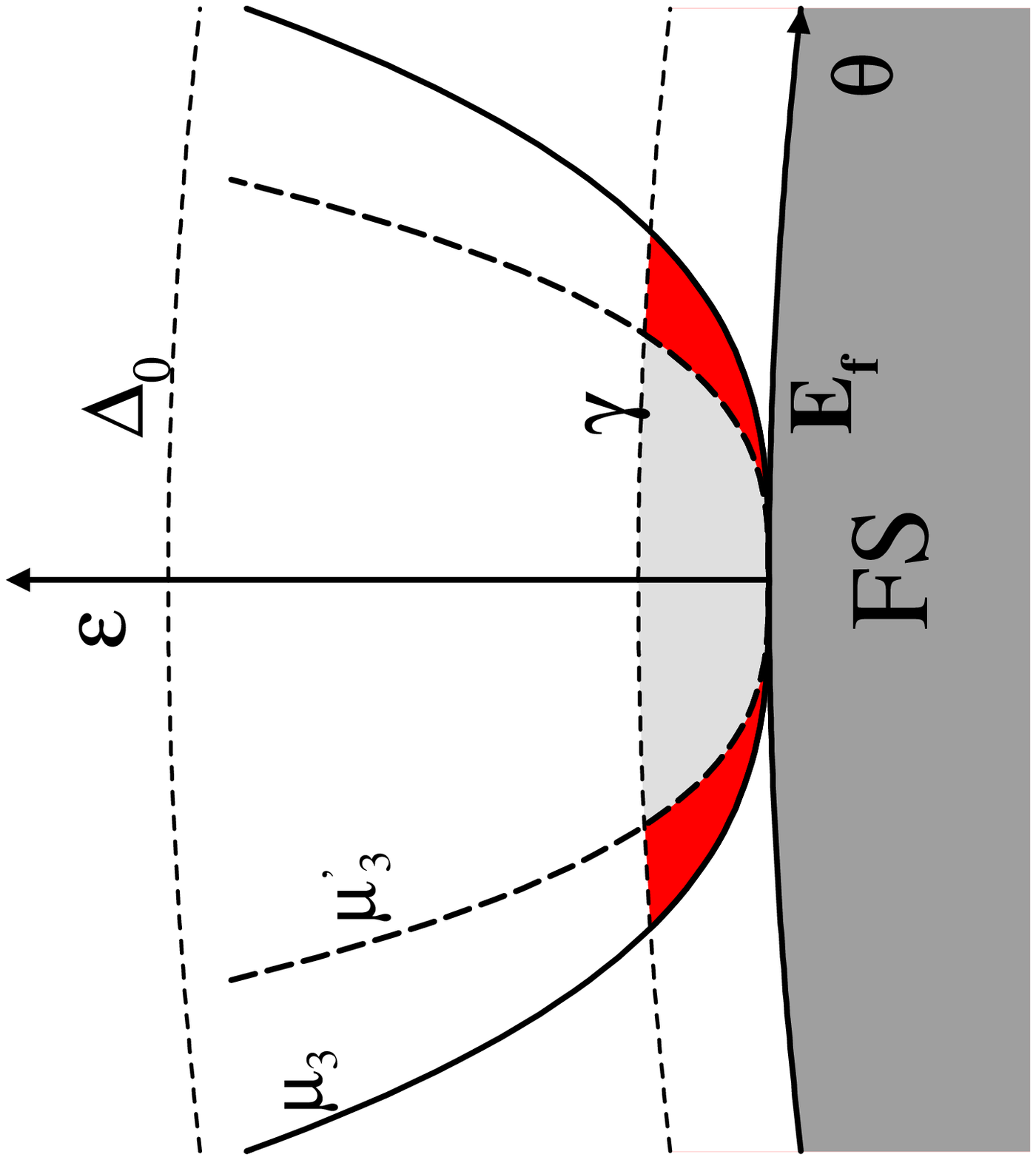}}}
\end{minipage}

\begin{center}
\begin{minipage}{1.0\textwidth}
{\small Fig.\ 2.
The gap function $|\Delta(\theta)| 
\approx \mu_3 \Delta_0|\theta|^3$ at the Fermi
surface (FS) near a cubic point node. At low temperatures $T\alt \gamma \ll 
\Delta_0$ the excitations are confined to a lower dimensional region of the 
Fermi surface, i.e. $0 \leq \epsilon \alt \gamma$ near the node.  
The solid (dashed) line corresponds to a gap function 
with nodal parameter $\mu_3$ ($\mu_3^\prime$).
The dark shaded area is a measure of the increase in
phase space for $\mu_3 < \mu_3^\prime$.}
\end{minipage}
\end{center}
\vspace{0.1in}

For $\mu_3\alt 1$ the low temperature limit of the anisotropy ratio of the
thermal conductivity for the B$_{1u}$ state is,\cite{gra96a}
\beq\label{ratio_B1u}
\cR\equiv\frac{ \kappa_c(0)/\kappa_b(0) }{ \kappa_c(T_c)/\kappa_b(T_c) }
\sim  10 \,\frac{ \mu }{ \mu_3 }
\left( \frac{\mu_3\Delta_0}{\gamma} \right)^{1/3} \,,
\eeq
where $\mu$ is the nodal parameter for the 
axial line nodes, i.e.
$\Delta(\phi,\theta\!=\!\pi/2) \approx \mu \Delta_0 \phi$.
For $\mu_3\sim 1$ and 
$\gamma/\Delta_0 \ll 1$ the ratio is typically much larger than unity.
Based on Eq.(\ref{ratio_B1u}) it appears 
possible to accomodate the experimental
constraint, $\cR_{\mbox{\small expt}} < 1$, 
with a B$_{1u}$ order parameter
by increasing $\mu_3$, i.e. decreasing
the phase space associated from 
the cubic point node.\cite{gra96a}
However, Eq. (\ref{ratio_B1u}) breaks down for $\mu_3 > 1$.
This corresponds to shrinking and effectively removing the cubic point
nodes.  In this limit the axial line nodes determine the transport
coefficients for both the $c$-axis and $ab$-plane currents. 
We find that $\cR\ge 1$, and in the limit $\mu_3\gg 1$ the anisotropy ratio
approaches unity (see also Table~2).

The result $\cR\ge 1$ also applies to the \AB-model.  
This can be understood by noting that
in the calculation of the thermal conductivity the order parameter
enters as $\Delta \Delta^*$. For the A$_{2u}\oplus$B$_{1u}$ 
order parameter the product $\Delta \Delta^* =
|\Delta^{\tiny A_{2u}}(\vp_{\!f}) + i \Delta^{\tiny B_{1u}}(\vp_{\!f})|^2$
can always be written in the form
$|\Delta^{\tiny B_{1u}}(\vp_{\!f})|^2(1+\cF(\vp_f))$, where $\cF$ is
an invariant function. This is precisely the
variational form for the product for the basis 
function with B$_{1u}$ symmetry
discussed above. Thus, we conclude that
the A$_{2u}\oplus$B$_{1u}$ model
cannot accomodate the experimental result 
$\cR_{\mbox{\small expt}} < 1$.
The same arguments can be applied to the
other relevant order parameter
models of the \AB-class, e.g. A$_{1u}\oplus$B$_{1u}$ and A$_{1g}\oplus$B$_{2g}$.
Thus, for a unitary order parameter belonging to the spin version of the
\ESG-models (e.g. B$_{1u}$), and for the same parity subclasses of the
\AB-models, e.g. A$_{2u}\oplus$B$_{1u})$, 
the anisotropy ratio is always greater than unity, which disagrees 
with the experimental results for $\cR$.\cite{lus96b}.

\vspace{0.3in}
{\small \noindent Table~2.\
Asymptotic values of the thermal conductivity tensor
{\boldmath$\kappa$}$/T$ for $T\to 0$.}
\begin{flushleft}
\renewcommand{\arraystretch}{1.2}
\begin{tabular}{cccc}
\noalign{\smallskip}\hline\hline
     Rep 
  &
  & {$\displaystyle {\kappa_b(T)}
	\left(\frac{1}{2}\gamma_S {T}\,{v_{\!f,b}^2} \right)^{-1}$}
  & {$\displaystyle {\kappa_c(T)}
	\left(\frac{1}{2}\gamma_S {T}\,{v_{\!f,c}^2} \right)^{-1}$}
\\ \hline
    $\rm E_{1g}$
  & \mbox{}\quad $\mu_1 \sim 1$
  & $\displaystyle {1}/({2\mu\Delta_0})$
  & $\displaystyle {\gamma}/({\mu_1^2\Delta_0^2})$
\\[1.0ex]
    $\rm E_{2u}$
  & \mbox{}\quad $\mu_2 \sim 1$
  & $\displaystyle {1}/({2\mu\Delta_0})$
  & $\displaystyle {1}/({2\mu_2\Delta_0})$
\\[1.0ex]
    $\rm B_{1u}$
  & \mbox{}\quad $\mu_3 \sim 1$
  & $\displaystyle {3}/({2\mu\Delta_0})$
  & $\displaystyle \frac{\sim 1}{\mu_3\Delta_0}
	\left(	{\mu_3\Delta_0}/{\gamma} \right)^{1\over 3}$
\\
  &
  \mbox{}\quad $\mu_3 \gg 1$
  & $\displaystyle {2}/({\pi\mu\Delta_0})$
  & $\displaystyle {2}/({\pi\mu\Delta_0})$
\\[1.0ex]
    $\rm A_{2u}\! \oplus B_{1u}$
  & \mbox{}\quad $\mu_3 \sim 1$
  & $\displaystyle {3}/({2\mu\Delta_0^{\tiny B_{1u}}})$
  & $\displaystyle \frac{\sim 1}{\mu_3\Delta_0^{\tiny B_{1u}}}
	\left(	{\mu_3\Delta_0^{\tiny B_{1u}}}/{\gamma} \right)^{1\over 3}$
\\
  &
  \mbox{}\quad $\mu_3 \gg 1$
  & $\displaystyle {2}/({\pi\mu\Delta_0^{\tiny B_{1u}})}$
  & $\displaystyle {2}/({\pi\mu\Delta_0^{\tiny B_{1u}})}$
\\[1.0ex]
    $\rm A_{1g}\! \oplus E_{1g}$
  & \mbox{}\quad $\mu_A, \mu_E \sim 1$
  & $\displaystyle \frac{\sim \gamma^3}{
    \left(\mu_A\Delta_0^{\tiny A}\right)
    \left(\mu_E\Delta_0^{\tiny E}\right)^3}$
  & $\displaystyle \frac{\sim \gamma}{
	\left(\mu_A\Delta_0^{\tiny A}\right)
	\left(\mu_E\Delta_0^{\tiny E}\right)}$
\\
\noalign{\smallskip}\hline\hline
\end{tabular}
\renewcommand{\arraystretch}{1.0}
\end{flushleft}
\vspace{0.1in}

Machida \et\cite{mac93,mac98} argue that UPt$_3$ is described
by a non-unitary spin-triplet ground state. Non-unitary
ground states break time-reversal symmetry in spin space; 
$\vS(\vp_f)=i\vDelta(\vp_f)\times\vDelta^*(\vp_f)$ 
is proportional to the spin polarization
of triplet pairs with relative
momenta $\vp_f$. The ground state of
the \ESG-model of Ref. (\onlinecite{mac93,mac98})
is a non-unitary state with
$\vDelta=\Delta(\vp_f)(\hat{\va}+i\hat{\vc})/2$.
The corresponding excitation spectrum has two branches:
E$_{\uparrow}=\sqrt{\xi_{\tiny\vp_f}^2+|\Delta(\vp_f)|^2 }$ and
E$_{\downarrow}=|\xi_{\tiny\vp_f}|$, 
where $\uparrow$ ($\downarrow$)
corresponds to an excitation with spin
parallel (anti-parallel) to $\vS$. Thus, the down-spin excitations
are gapless over the whole Fermi surface and give rise to a large 
heat current at low temperatures compared to unitary or even parity
superconducting states. The low temperature thermal conductivity 
is greater than half the normal-state value
extrapolated below $T_c$, i.e. $\kappa/T\ge\frac{1}{2}\,\kappa_N/T$.
From Ref. (\citen{lus96b}) we deduce that the normal-state
scattering rate extrapolated to $T=0$ is $1/\tau(0)\approx\frac{1}{2}
\times 1/\tau(T_c)$. This gives $\lim_{T\to 0} \kappa/T \agt [\kappa_N/T]_{T_c}$,
a result which is in striking contrast with the low temperature experimental 
data for UPt$_3$, which are roughly two orders of magnitude smaller, i.e.
$\lim_{T\to 0} \kappa_i/T \sim 0.01 [\kappa_i/T]_{T_c}$, with $i=a,c$.
Thus, a non-unitary ground state of the form proposed
by Machida \et\cite{mac93,mac98} is incompatible with the
thermal conductivity data on UPt$_3$.\footnote{
Early transport and heat capacity experiments at higher temperatures
on poorer quality crystals were interpreted in terms of a large residual
density of states at the Fermi energy.  Indeed this interpretation was
the original argument for introducing the non-unitary spin-triplet pairing
model for UPt$_3$.}

Now consider the orbital \ESG-model proposed by Zhitomirsky and
Ueda.\cite{zhi96}
They start from the five-dimensional $\ell=2$ orbital representation
of the full rotation group and assume that crystal-field terms
break the full rotational symmetry and split
the d-wave reprentation into three irreducible representations
of $D_{6h}$; $\Gamma_2[SO(3)]\Rightarrow$ A$_{1g}\oplus$E$_{1g}\oplus$E$_{2g}$
with $T_{c}^{A_{1g}} > T_{c}^{E_{1g}} \gg T_{c}^{E_{2g}}$.
The onset of superconductivity is identified with the
A$_{1g}$ component, and the second transition is identified with
nucleation of an E$_{1g}$ component. The \AE-model would in principle
predict a third transition involving the E$_{2g}$ amplitude,
however, no such transition is observed in UPt$_3$. Thus, $T_c^{E_{2g}}$
must be suppressed to very low temperatures by the development
of the first two order parameter components.

The onset of an E$_{1g}$ order parameter spontaneously 
breaks both time-reversal symmetry and rotational symmetry in the 
basal plane. The ground-state order parameter in the \AE-model
has the form,
$\Delta^{AE}({\bf p}_f) = \Delta^{A_{1g}}({\bf p}_f) + i
\Delta^{E_{1g}}({\bf p}_f)$, with possible basis functions
\ber
\Delta^{A_{1g}}({\bf p}_f) &=&
  \left[ 
   \Delta_0^A(T) \frac{\tiny 1}{\tiny 2}( 3\cos(\theta)^2 - 1 )
  \right]
  {\cal F}_{A_{1g}}({\bf p}_f; \mu_A)\,,
\\
\Delta^{E_{1g}}({\bf p}_f) &=&
  \left[ 
   \Delta_0^E(T)  \sin(2\theta) \sin(\phi)
  \right]
  {\cal F}_{E_{1g}}({\bf p}_f; \mu_E) \, .
\eer
The A$_{1g}$ basis function has tropical line nodes above and
below the equator at $\theta=\cos^{-1}(\pm 1/\sqrt{3})$.
The E$_{1g}$ basis function has an axial
line node running from the north pole to the south pole.  Consequently the
ground-state \AE~ order parameter
has four cross nodes located at the intersection of the 
tropical and axial line nodes.
As can be seen from Table~2 the asymptotic values
of the thermal conductivities of the \AE-model are nonuniversal.\footnote{ 
An extra complication in the analysis of the \AE-model
is that the angular average of the order parameter
is (generally) nonvanishing, 
$\langle \Delta^{A_{1g}}({\bf p}_f) \rangle \neq 0$,
which results in non-vanishing off-diagonal contributions to the 
scattering self-energy which must be calculated self-consistently.}

An additional feature of the \AE-model is that
the low-temperature phase spontaneously
breaks rotational symmetry in the $ab$-plane. The low-temperature
phase of the Ginzburg-Landau (GL) functional for the 
\AE-model is also degenerate with
respect to rotations of the E$_{1g}$ order parameter in the $ab$-plane.
This accidental degeneracy is lifted by higher order terms in the
free energy (sixth-order and higher) which differentiate
hexagonal symmetry from cylindrical symmetry.
In order for the \AE-model to account for the 
thermal conductivity data for heat flow along the $c$ and $b$ axes
the ground-state must {\it orient} such 
that the axial line node of the E$_{1g}$ order parameter is
along the $a$-direction ($k_x$-direction).
If the hexagonal anisotropy energy is small, $\cF_{hex}\ll N_f
|\Delta|^2$, the heat current may orient the E$_{1g}$ order parameter
so that the current response is nearly isotropic in the $ab$-plane. However,
for sufficiently small currents the anisotropy of the thermal conductivity
tensor that results from the spontaneously broken 
rotational symmetry in the $ab$-plane should be observable.
Table~2 summarizes the low temperature limits for the 
thermal conductitivities for the models discussed above.

\section*{Numerical Results and Conclusions}

In order to obtain quantitative results for the thermal conductivity over
the full temperature range below $T_c$ we must include inelastic
`electron-electron' scattering.
We model the inelastic scattering rate in the 
superconducting phase of UPt$_3$ by extrapolating the normal-state
scattering rate, $1/\tau_{\small ee}=A T^2$, below $T_c$. For $T\alt0.3T_c$
the inelastic channel is unimportant even for the clean limit of
UPt$_3$ discussed here. The combined elastic and inelastic
scattering rate in the normal-state is $\Gamma(T)=\Gamma_0+A\,T^2$.
From the measurements in
Ref.~\citen{lus96b} it follows that the UPt$_3$ crystals that were studied
have nearly equal elastic and inelastic scattering rates at $T_c$, i.e.
$\Gamma_0=A\,T_c^2$, so that $\Gamma(T) = \Gamma_0 (1+T^2/T_c^2)$.
Our procedure for computing and fitting the thermal conductivities 
is described in detail in Refs.~\onlinecite{gra96a} and \onlinecite{gra96b}.
Below $T_c$ we assume that the elastic scattering 
is in the unitarity limit.\cite{gra96b}
Here we briefly outline the procedure and then discuss the numerical 
results for the B$_{1u}$, \AB, and \AE~ models.
  
We describe the analysis of the B$_{1u}$ order parameter; a similar analysis
is carried out for the other order parameter models.
We first calculate the basal plane thermal conductivity, $\kappa_b$,
and adjust the scattering rate $\Gamma_0$ and the nodal parameters, 
$\mu$ at $\phi=0, \theta=\pi/2$ and 
$\mu_3$ at $\phi=\pi/6, \theta=0$, to obtain a good fit between theory 
and experiment. The next step is to compute $\kappa_b$ and $\kappa_c$ keeping 
$\Gamma_0$ and $\mu$ fixed, and relaxing $\mu_3$ for a best fit.
Figure~3 shows the nearly perfect fit to the basal plane thermal
conductivity of Ref. (\onlinecite{lus96b}) for the B$_{1u}$ model.

For the variational function ${\cal F}_{\tiny B_{1u}}$ we have 
chosen the form
\beq
{\cal F}_{\tiny B_{1u}} = y_0(a_\phi,a_\theta) 
\left(
  1 + a_\phi \, \cos^2(3\phi) + a_\theta \, \cos^2(\theta) 
\right) \,,
\eeq
where the prefactor $y_0$ is determined by 
${\rm max}\, \Delta(\vp_f) = \Delta_0$.
The coefficients $(a_{\phi},a_{\theta})$ determine the nodal
parameters,
\ber
\mu &=& 3(1 + a_\phi)\, y_0(a_\phi,a_\theta) \,, \\
\mu_{3} 	 &=& (1 + a_\theta)\, y_0(a_\phi,a_\theta) \,.
\eer
Thus, we have two variational parameters $(a_{\phi},a_{\theta})$ 
that enable us to
independently adjust the nodal parameters 
$\mu$ and $\mu_{3}$.

We also examined the sensitivity of our numerical results to the form
of the variational function. In particular, we calculated the thermal
conductivity for a piecewise continuous B$_{1u}$ order parameter of the form
$\Delta(\phi,\theta) = \Delta_0^{\tiny B_{1u}} g(\phi) h(\theta)$ with
\ber
g(\phi) &=& \left\{
  \begin{array}{r@{\quad:\quad}l}
	\mu \Delta_0 \phi & 0 \le \phi < \phi_* \,,
\\
   1 & \phi_* \le \phi \le \frac{\pi}{6}\,,
  \end{array}
\right.
\\
h(\theta) &=& \left\{
  \begin{array}{r@{\quad:\quad}l}
	\mu_{3} \Delta_0 \theta^3 & 0 \le \theta < \theta_* \,,
\\
   1 & \theta_* \le \theta \le \frac{\pi}{2}\,,
  \end{array}
\right.
\eer
where
$\phi_* = {\rm min}( \frac{\pi}{6}, {1}/{\mu} \Delta_0 )$, and
$\theta_* = {\rm min}( \frac{\pi}{2}, {1}/{(\mu_{3} \Delta_0)^{1/3}} )$, 
and the imposed B$_{1u}$ symmetry on the functions $g(\phi)h(\theta)$; i.e.
$g({\tiny \pi\over 3}-\phi) = g(\phi)$, 
$g({\tiny \pi\over 3}+\phi) = - g(\phi)$, 
and $h(\pi-\theta) = h(\theta)$.

\vspace{0.1in}
\noindent
\begin{minipage}{0.491\textwidth}
\centerline{ \epsfxsize=\hsize{ \epsfbox{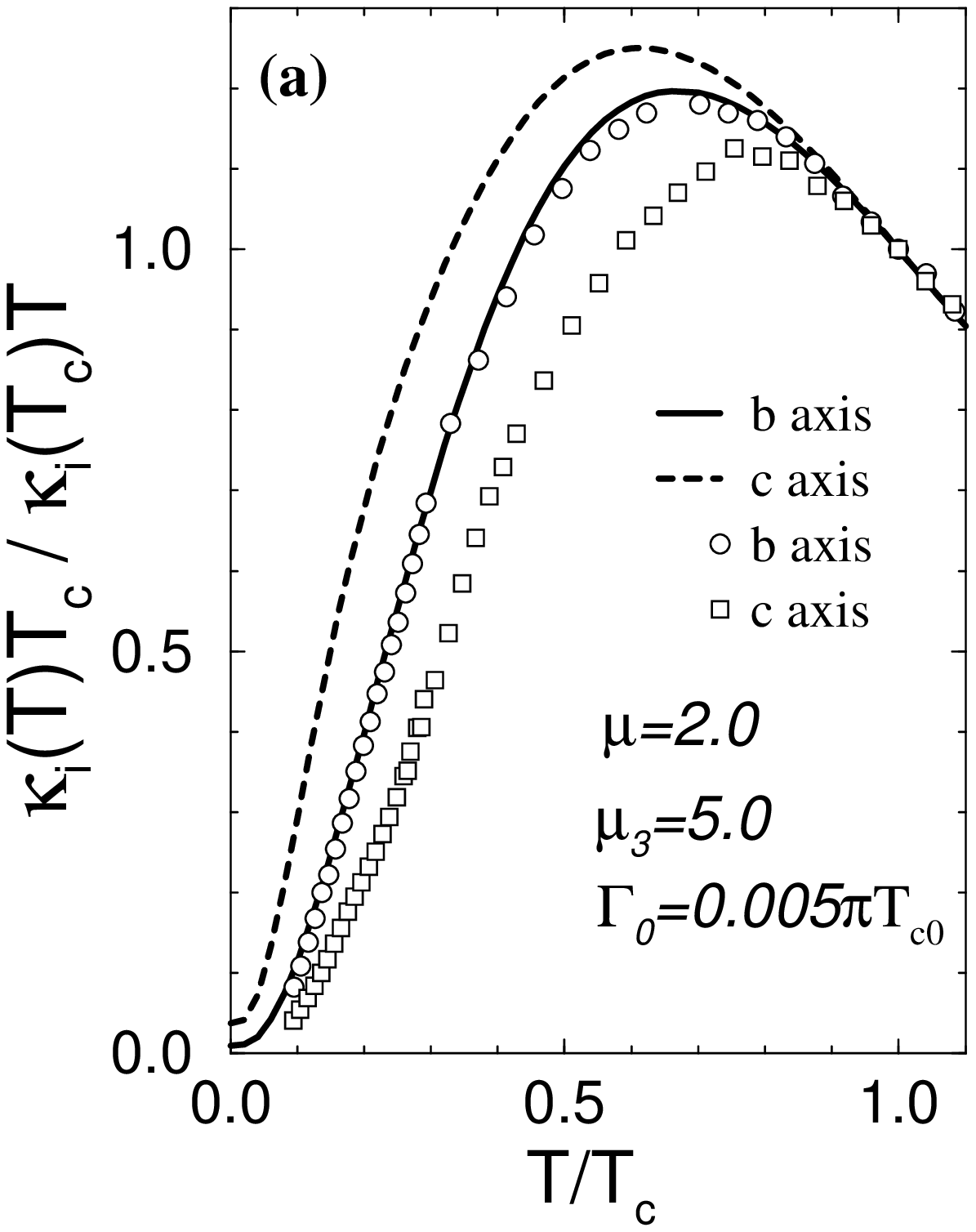} } }
\end{minipage}
\hfill
\begin{minipage}{0.491\textwidth}
\centerline{ \epsfxsize=\hsize{ \epsfbox{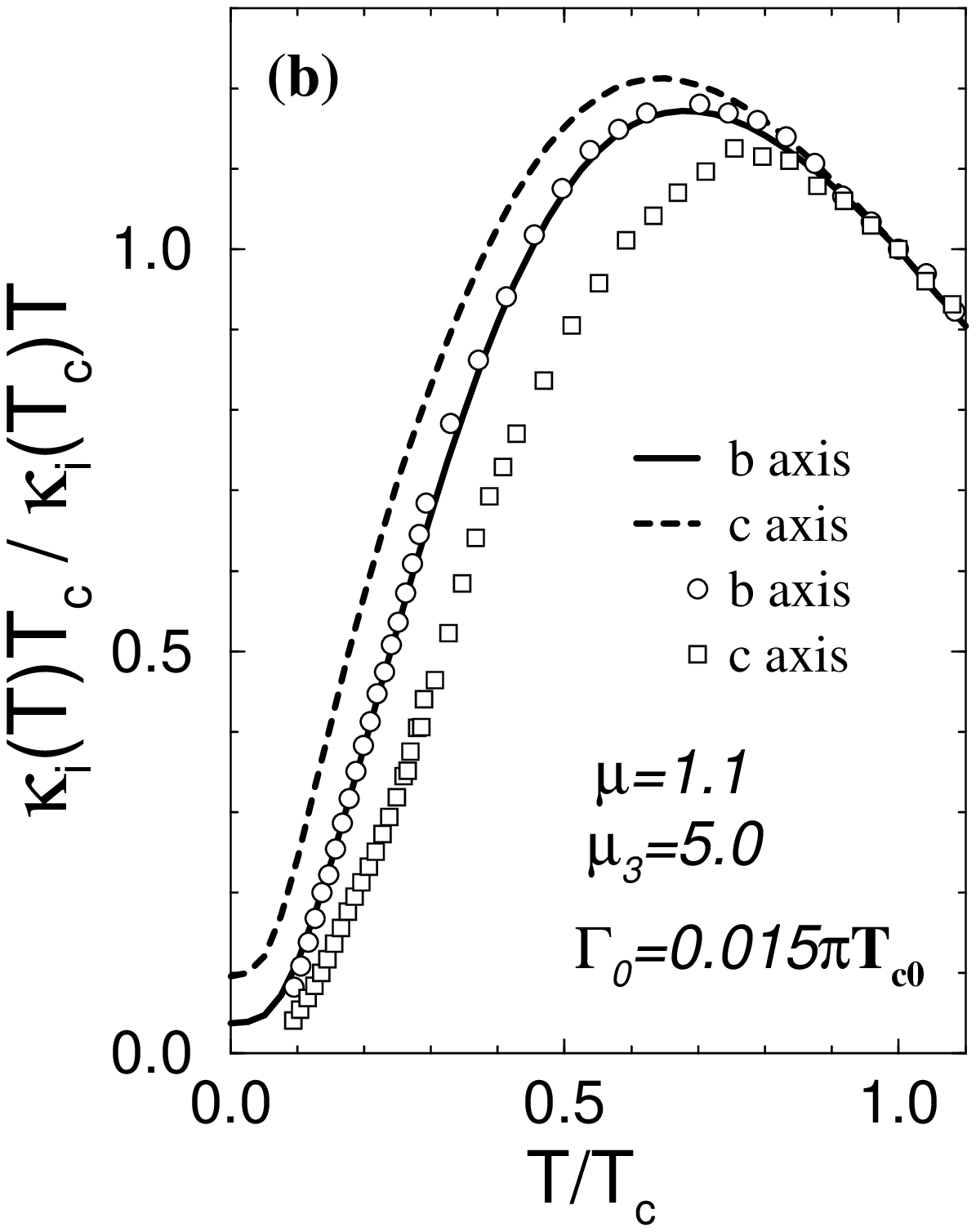} } }
\end{minipage}
\noindent
\begin{minipage}{\textwidth}
\centerline{ \epsfysize=0.9\hsize 
             \rotate[r]{ \epsfbox{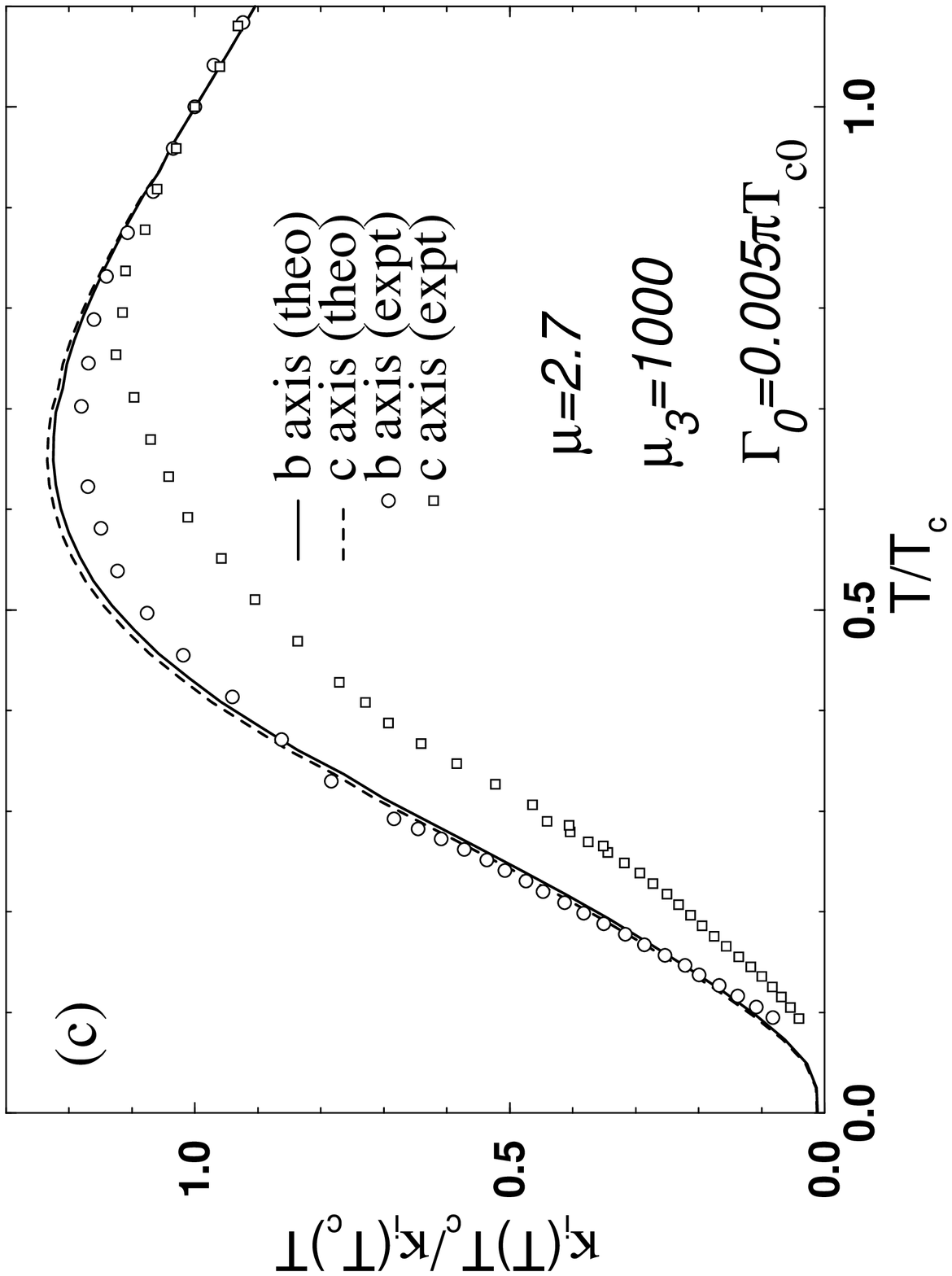} } }
\end{minipage}

\begin{center}
\begin{minipage}{1.0\textwidth}
{\small Fig.\ 3. (a) The normalized thermal conductivities for a B$_{1u}$ state
in comparison with the experimental data on UPt$_3$.\cite{lus96b}} The best
fit parameters are shown in the figure. (b) The same comparison for the
A$_{2u}\oplus B_{1u}$ model, and (c) for a piecewise continuous B$_{1u}$
state (see text).
\end{minipage}
\end{center}
\newpage

With this piecewise B$_{1u}$ model we confirm our earlier
conclusions, {\it viz.}  that only the
values of $\Delta_0$, $\Gamma_0$, and $\{\mu_i\}$ vary slightly,
but the best fit low-temperature thermal response functions are unaffected.
Furthermore, in the limit $\mu_3 \gg 1$ (ie. $h(\theta) \to 1$)
we obtain $\cR\rightarrow 1$ for the asymptotic anisotropy ratio
(see Fig.~3c).
In short, all the numerical results are in agreement with our estimates of
the asymptotic values of the B$_{1u}$ pair state. These considerations
apply to the analysis of the \AB~ and \ESG-models as well.

Figure~3 shows our `best fit' calculations for the B$_{1u}$ (Fig. 3a)
and \AB~(Fig. 3b) order parameter models. For both models we obtain
excellent fits to the basal plane thermal conductivity, $\kappa_b/T$,
over the entire temperature range. However, we are unable to fit the
$c$-axis thermal conductivity with the B$_{1u}$ and \AB~ 
order parameter models, which is consistent with our phase space
arguments and asymptotic estimates.

We obtain good low temperature fits for both the $b$ and $c$
axis thermal conductivity with the \AE-model provided the E$_{1g}$
component is oriented with the axial line node along the $a$-axis.
The insert in Fig.~4 shows the strong anisotropy of the heat
flow in the $ab$-plane for the \AE-model, \ie
$\lim_{T\rightarrow 0}\kappa_a/\kappa_b \sim 4$, which is
a direct consequence of the broken rotational symmetry in 
the basal plane. This anisotropy, which is specific to the
the \AE-model, has not been experimentally observed. It would
serve as an important test of the \AE-model and would differentiate
the model from the {\it 2D} E-rep models. The nodal
parameters, which determine the spectrum of low-lying
excitations around the cross nodes, are defined as
$\Delta^{A_{1g}}(\theta) \approx \Delta_0^A \mu_A (\theta_* - \theta)$,
with $\theta_* = \cos^{-1} \sqrt{1/3}$,
and 
$\Delta^{E_{1g}}(\phi) \approx \Delta_0^E \mu_E \phi$.

The quality of our fits for the thermal conductivity along different
crystal axes is best summarized in terms of the anisotropy ratios
$\cR(T)=[\kappa_c(T)/\kappa_b(T)]/(\kappa_c(T_c)/\kappa_b(T_c)]$ for
the various order parameter models. Figure~5 shows that
the anisotropy ratio for the data of Ref. (\onlinecite{lus96b})
is well described by the ground states for the
E$_{1g}$, E$_{2u}$ and A$_{1g}\oplus$E$_{1g}$ models.
The \AB-models, and the spin-triplet \ESG-models
cannot account for the anisotropy of the
thermal conductivity at low temperatures.

Independent measurements of the thermal conductivity 
by Suderow \et\cite{sud97} differ somewhat from 
those of Lussier {\it et al.},\cite{lus96b} particularly
in the anisotropy ratio ${\cal R}$.
This difference is mainly due to differences in the
$c$-axis thermal conductivity. The normalized
basal plane conductivities, plotted vs. $T/T_c$, are in remarkably
good agreement, despite of being measured on different samples.
The differences in the data for heat flow along the $c$-axis 
for the different samples might be due to crystal imperfections along 
the $c$-axis, such as stacking faults, structural modulations or a 
mosaic structure.

\noindent
\begin{minipage}{\textwidth}
\centerline{ \epsfysize=0.84\hsize \rotate[r]{ \epsfbox{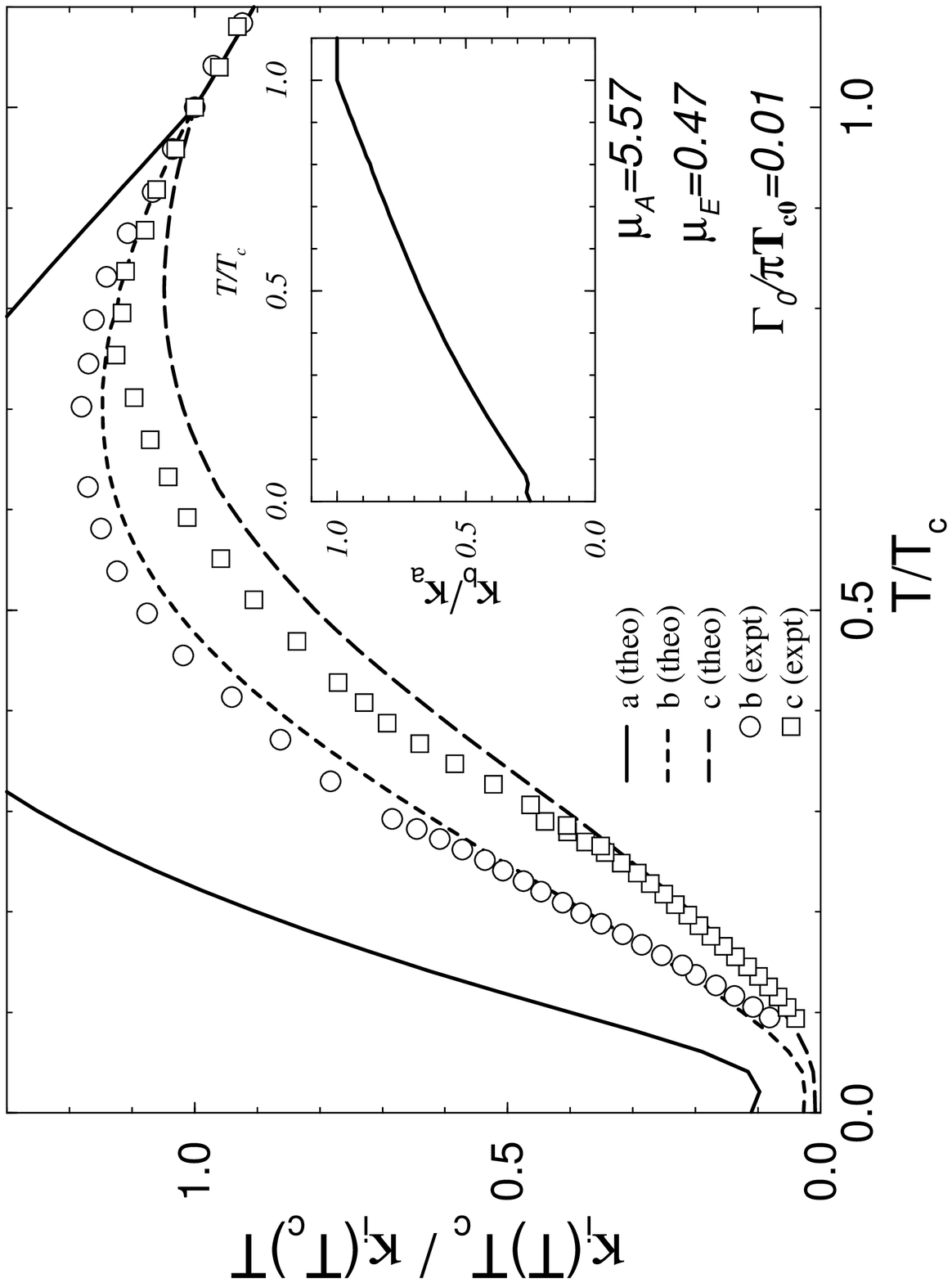} } }
\end{minipage}

\begin{center}
\begin{minipage}{1.0\textwidth}
{\small Fig.\ 4.
The normalized thermal conductivities for the A$_{1g}\oplus$E$_{1g}$ model
in comparison with the experimental data on UPt$_3$.\cite{lus96b}
The inset shows the ratio of the conductivities $\kappa_b/\kappa_a$.
}
\end{minipage}
\end{center}

\noindent
\begin{minipage}{\textwidth}
\centerline{ \epsfysize=0.95\hsize\rotate[r]{ \epsfbox{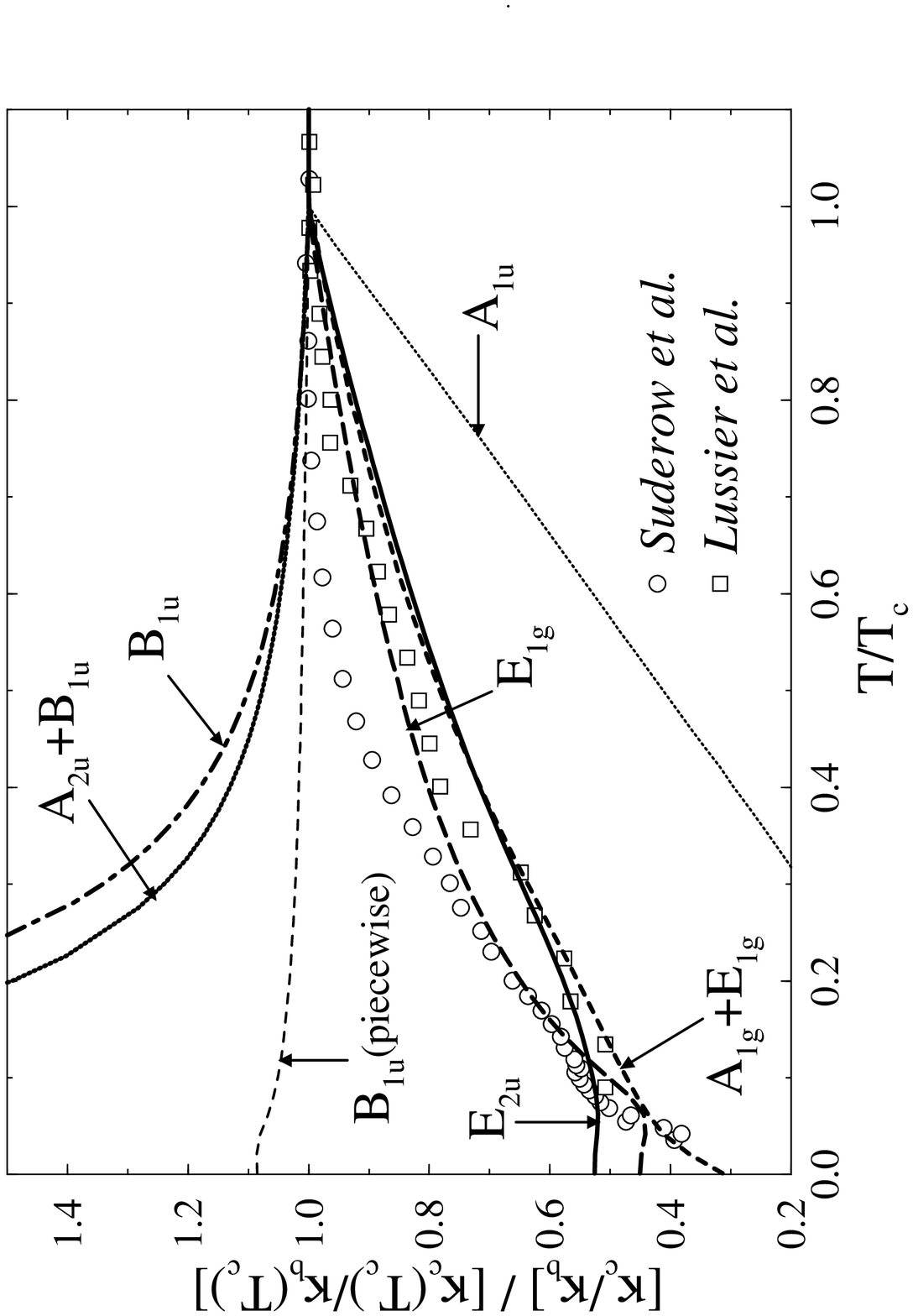} } }
\end{minipage}

\begin{center}
\begin{minipage}{1.0\textwidth}
{\small Fig.\ 5. The anisotropy ratio of the thermal conductivities for
various `best fit' pairing states. The lines for an E$_{1g}$, E$_{2u}$
and A$_{1u}$ state are shown for comparison.}
\end{minipage}
\end{center}
\vspace{0.1in}

\section*{Conclusion}

Our analysis of the low temperature thermal conductivities for a single
component order parameter of the \ESG-class (\eg B$_{1u}$), and the
accidental degeneracy order parameter models of the \AB-class
(\eg A$_{2u}\oplus$B$_{1u}$), and the orbital subclass of the \ESG-class
(A$_{1g}\oplus$E$_{1g}$), support earlier analyses suggesting that one of 
the {\it 2D} E-rep models is a promising model for the order 
parameter of UPt$_3$.
The \AB-models and the spin-triplet version of the \ESG-model are in
substantial disagreement with the experimental data at low temperatures.
Cubic point nodes or cross nodes at or near the
poles of the Fermi surface lead to a large heat current along the $c$-axis, 
which is not reconcilable with the experimental data.
The two-component orbital \AE-model can account for the $c$-axis and $b$-axis
thermal conductivities with an appropriate orientation of the E$_{1g}$
order parameter. However, the \AE-model spontaneously breaks rotational
symmetry in the basal plane which so far has not been observed in any experiment
to date. The basal plane anisotropy is large ($\lim_{T\rightarrow 0}
\kappa_a/\kappa_b\sim 4$) and should easily be detected.

\section*{Acknowledgments}
This research was supported by the National Science Foundation
(grant no. DMR-9705473) and the Science and Technology Center for
Superconductivity (grant no. DMR 91-20000).  MJG also acknowledges support
from Los Alamos National Laboratory under the auspices of the U.S.\
Department of Energy, where this work was completed.


\end{document}